\documentclass[sigconf]{acmart}

\AtBeginDocument{%
  }

\setcopyright{acmlicensed}
\copyrightyear{2018}
\acmYear{2018}
\acmDOI{XXXXXXX.XXXXXXX}

\acmConference[Conference acronym 'XX]{Make sure to enter the correct
  conference title from your rights confirmation emai}{June 03--05,
  2018}{Woodstock, NY}
\acmISBN{978-1-4503-XXXX-X/18/06}

\usepackage{multirow}
\usepackage{marvosym}
\usepackage{diagbox}
\usepackage{graphicx}
\usepackage{subcaption}
\usepackage{balance}
\begin{document}

\title{FishBargain: An LLM-Empowered Bargaining Agent for Online Fleamarket Platform Sellers}


\author{Dexin Kong}
% \authornote{Corresponding Author}
\email{kongdexin.kdx@taobao.com}
\affiliation{%
  \institution{Alibaba Group}
  \city{Hangzhou}
  \country{China}}
  
\author{Xu Yan}
% \authornote{Corresponding Author}
\email{wuyong.yx@taobao.com}
\affiliation{%
  \institution{Alibaba Group}
  \city{Hangzhou}
  \country{China}}

\author{Ming Chen}
% \authornote{Corresponding Author}
\email{xingke.cm@taobao.com}
\affiliation{%
  \institution{Alibaba Group}
  \city{Hangzhou}
  \country{China}}
  
\author{Shuguang Han\textsuperscript{\Letter}}
% \authornote{Corresponding Author}
\email{shuguang.sh@alibaba-inc.com}
\affiliation{%
  \institution{Alibaba Group}
  \city{Hangzhou}
  \country{China}}

\author{Jufeng Chen}
\email{jufeng.cjf@alibaba-inc.com}
\affiliation{%
  \institution{Alibaba Group}
  \city{Hangzhou}
  \country{China}}
  
\author{Fei Huang}
\email{huangfei.hf@taobao.com}
\affiliation{%
  \institution{Alibaba Group}
  \city{Hangzhou}
  \country{China}}
% \author{Jufeng Chen}
% \email{jufeng.cjf@alibaba-inc.com}
% \affiliation{%
%   \institution{Alibaba Group}
%   \city{Hangzhou}

\renewcommand{\shortauthors}{Kong et al.}
\newcommand{\runzhe}[1]{\textcolor{cyan}{}}

\begin{abstract}
Different from traditional Business-to-Consumer e-commerce platforms~(e.g., Amazon), online fleamarket platforms~(e.g., Craigslist) mainly focus on individual sellers who are lack of time investment and business proficiency. Individual sellers often struggle with the bargaining process and thus the deal is unaccomplished. Recent advancements in Large Language Models(LLMs) demonstrate huge potential in various dialogue tasks, but those tasks are mainly in the form of passively following user's instruction. Bargaining, as a form of proactive dialogue task, represents a distinct art of dialogue considering the dynamism of environment and uncertainty of adversary strategies. In this paper, we propose an LLM-empowered bargaining agent designed for online fleamarket platform sellers, named as FishBargain. Specifically, FishBargain understands the chat context and product information, chooses both action and language skill considering possible adversary actions and generates utterances. FishBargain has been tested by thousands of individual sellers on one of the largest online fleamarket platforms~(Xianyu) in China. Both qualitative and quantitative experiments demonstrate that FishBargain can effectively help sellers make more deals. 
\end{abstract}





\maketitle

\section{INTRODUCTION}

\begin{figure}[ht]
  \centering
  % \captionsetup{skip=5pt}
  \includegraphics[width=0.4\textwidth]{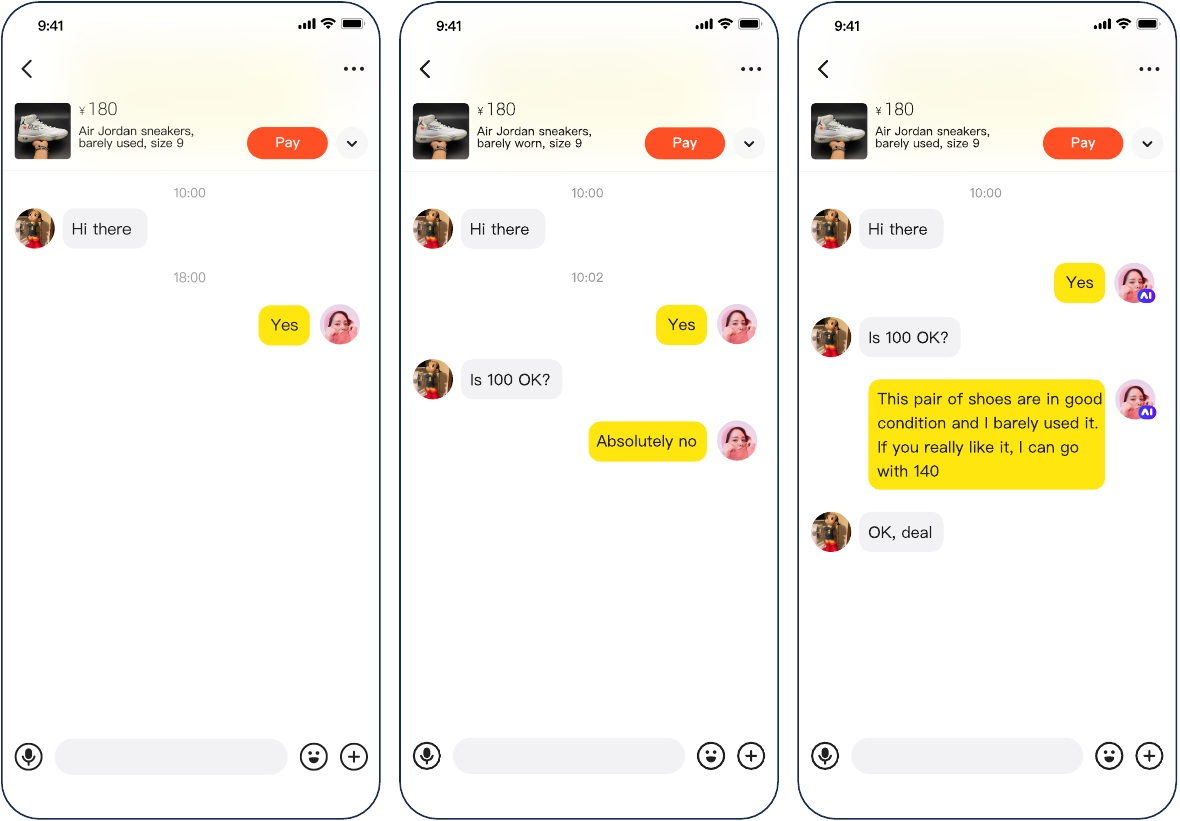}
  % \vspace{-5pt}
  \caption{Bargaining process on the Xianyu platform. The two on the left are human responses. Failure to respond in time and lack of bargaining skills lead to a failed sale. The one on the right is the FishBargain response, which responds in time and closes the deal.}
  \label{fig:intro}
  % \vspace{-10pt}
\end{figure}

With the development of the economic society, online fleamarket platforms have grown rapidly around the world, allowing individual users to sell products online~\cite{DBLP:conf/emnlp/ChenZLJLHW0C24}. Products listed on online fleamarket platforms are in different conditions, ranging from brand-new to heavily-used, so there is no standard to guide how they should be priced. Due to that, during a typical fleamarket shopping process, bargaining is almost an unavoidable part, which is different from business-to-consumer e-commerce platforms~(e.g., Amazon). Individual sellers often struggle with the bargaining process for two main reasons, as shown in Figure~\ref{fig:intro}. First, the time they invest on the platforms is limited, so when an interested buyer reaches out, the seller is often not available. We named it buyer-seller availability gap. Second, even if the seller is online, their insufficiency in negotiation techniques often leads to an unsatisfactory outcome. 
\begin{figure*}[h]
    \centering  
    \includegraphics[width=0.9\textwidth]{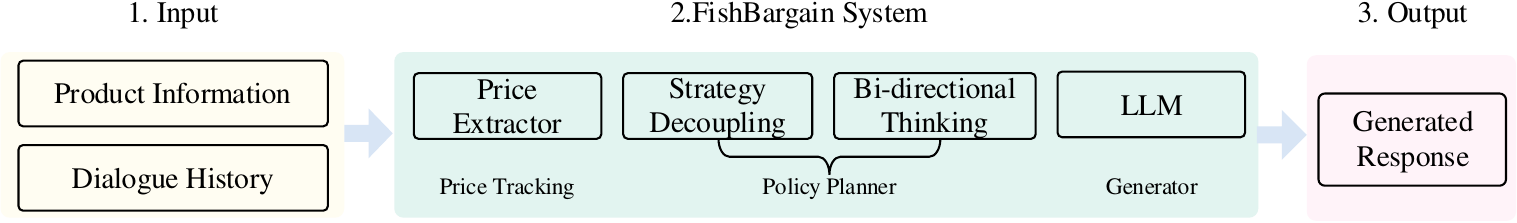}
    % \vspace{-5pt}
    \caption{Overview of the FishBargain system architecture.}
    \label{fig:sys}
    % \vspace{-10pt}
\end{figure*}

To address the issues mentioned above, it is important to introduce bargaining agent to help individual sellers. In recent years, LLM-driven agents have shown huge potential in various dialogue tasks, such as context understanding and response generation~\cite{ DBLP:conf/emnlp/ZhangPLZM23}. However, those tasks are mainly targeted at passively following users' intentions and fulfilling users' requests. Unlike those tasks, proactive dialogue tasks constitute a unique form of communication that takes into account environment and unpredictable opponent strategies~\cite{DBLP:conf/ijcai/0002LLC23}. Bargaining on fleamarket platforms is naturally one of the proactive dialogue tasks. In a typical fleamarket bargaining scenario, buyer and seller have conflict of interests~(target price) and seek a mutual agreement through dialogues. Such dialogues consist of cooperative and adversarial elements, where both sides need to understand the state, choose actions according to certain strategies and generate utterances to interact with each other~\cite{DBLP:conf/lrec/AsherHMBA16}. The whole process is both a strategic and a linguistic problem.

To this end, we propose an LLM-empowered bargaining agent named FishBargain, which consists of three modules: price extractor, policy planner and utterance generator, aiming to automate and improve the bargaining process on fleamarket platforms for individual sellers. Our contributions are threefold:

First, directly applying LLMs to generate responses often results in degenerate solutions because it is challenging to control and interpret the strategies. Preliminary investigations have attempted to address this by decoupling strategy selection from utterance generation~\cite{DBLP:conf/emnlp/HeCBL18}. However, generated utterances tend to be repetitive if the same action is chosen. Therefore, we take a step forward to divide strategies into both actions and language skills and introduce a policy planner to choose strategies. This methodology allows for a more nuanced and enriched expression in the generated utterances.
% the price is the key factor in bargaining, but recent studies demonstrate that LLMs are less capable with numbers~\cite{DBLP:journals/corr/abs-2410-05229}. To tackle this problem, we introduce a price extractor to track buyer's offers.

Second, rather than passively responding to the buyers, bargaining agents should have proactivity, defined as the capability to control the conversation considering the adversary actions and goals. So instead of choosing strategies only based on chat history, the policy planner is designed to conduct bi-directional thinking~\cite{DBLP:journals/corr/abs-2407-06112}, by anticipating the adversary's strategies.

Third, we define solid metrics and involve real sellers from Xianyu, to test our bargaining agent named FishBargain. 
% Since Xianyu means a species of fish in China, so we name the agent as . 
Both qualitative and quantitative evaluations demonstrate that FishBargain can effectively help individual sellers make more deals.

\section{OVERVIEW OF FISHBARGAIN SYSTEM}

Our goal is to create an agent to help sellers sell products at the highest possible price. For a given dialogue history $u_{<=t}$ and the product information $I$, this agent is responsible for generating the response $u_t^{sys}$. The agent is composed of the following three parts:
\begin{itemize}
\item A \textbf{price extractor}, which is responsible for extracting the price from the dialogue history $u_{<=t}$. 
% and giving the price in the response $u_t$.
\item A \textbf{policy planner} that predicts the next dialogue action $a_t$ and language skill $s_t$ based on the dialogue history and product information.
\item A \textbf{generator} that converts dialogue strategies and prices into natural language response $u_t^{sys}$.
\end{itemize}

\subsection{Price Extractor}

Price is a fundamental element in the bargaining process, but it is challenging for LLMs to interpret this factor. So we introduce a price extractor to alleviate the problem and keep track of the dynamics of buyer's offers. In the bargaining process, price can be expressed through absolute ways (e.g., "Is \$200 OK?") and relative ways (e.g., "How about a 20\% discount?"). To address the problem, the price extractor is required to extract the price according to chat history and also provide a detailed computation chain of thought to ensure the precision. 

Specifically, in a chat session concerning the product $I$, we denote each utterance as $u_t$, and the price extractor recognizes the buyer's price $p_t^{usr}$ by $f(I, u_{<=t})$. It is worth mentioning that the price extractor mainly focuses on the buyer side, because the seller's offer $p_t^{sys}$ is proposed by our system. 

\subsection{Policy Planner}

Policy planner decides what action $a_t$ the bargaining agent should take given the product $I$, utterance sequence $u_{<=t}$, and both sides' offer $p_{<=t}$. Below, we describe two key features of this module. 

\subsubsection{Strategy decoupling}
\begin{table}[hbp]
  \centering
  \caption{Our redesigned dialogue actions.}
  \vspace{-5pt}
  \label{tab:act}
    \begin{tabular}{lp{140pt}}
    \toprule
    Action & Definition \\
    \midrule
    \midrule
    DEAL  & Buyer and seller reach a deal \\
    PROPOSE & Initiate a price or a price range for the product. \\
    COUNTER & Propose a new price or a new price range. \\
    COUNTER-NOPRICE & Want to propose a new price but do not specifically mention a new price. \\
    % ABNORMAL & The buyer's bid is abnormal, such as higher than the product's price or lower than the price limit. \\
    REJECT & There is no room for negotiation. Stick to the bottom price. \\
    HELLO & Say hello or chat randomly. \\
    ANS   & Answer buyer questions based on the product information. \\
    % UNKNOWN & Unable to answer user questions based on existing product information \\
    % BYE   & Say goodbye to the user in a friendly way and welcome them to visit next time \\
    % RECONFIRM & Unable to understand the buyer's question and need to further inquire the buyer to understand the problem. \\
    \bottomrule
    \end{tabular}%
  \label{tab:addlabel}%
    % \vspace{-10pt}
\end{table}%
\begin{table*}[htbp]
  \centering
  \caption{Our designed bargaining language skills.}
  \vspace{-5pt}
  \label{tab:script}
    \begin{tabular}{ll}
    \toprule
    Language Skill & Definition \\
    \midrule
    \midrule
    Emphasis & Highlight the cost value, quality or bottom price of the product to show the rationality of the pricing. \\
    % Emphasis on quality & Highlight the high quality of the product and emphasize that its performance and durability exceed similar products. \\
    % Emphasis on the bottom line & Inform that the current price is the lowest price the merchant can offer and cannot be reduced any further, showing the rationality and sincerity of the pricing. \\
    Added Value & Provide additional value beyond the product, such as gifts, free shipping, etc. \\
    Emotional Strategy & Use humor, expressions, complaints, and identity recognition to resonate with the other party. \\
    Compare the Market & Compare the product with other products on the market to highlight the advantages of its own products. \\
    Transaction Guarantee & Promise to ensure transaction security and reliability by offering good after-sales service. \\
    Create Urgency & Create urgency by reminding that the product may sell out soon or prices may rise shortly. \\
    % Other bargaining language skills & Other bargaining techniques are used. \\
    Chat  & Do not use techniques and simply reply to the other party. \\
    \bottomrule
    \end{tabular}%
  \label{tab:addlabel}%
    % \vspace{-10pt}
\end{table*}%
LLMs have demonstrated excellent contextual understanding abilities across various dialogue tasks. However, most dialogues are still primarily human-led. In other words, it is challenging for LLMs to strategically guide conversations toward the desired objectives, such as closing a deal at a higher price. To enable LLMs to proactively guide dialogues towards certain goals, we reference previous work by defining and executing bargaining actions to steer the direction of the conversation, with specific bargaining actions detailed in Table~\ref{tab:act}. We also utilize an LLM as the strategy planner to select the next action to execute based on the context.
$$a_t = LLM(P_a; I; u_1^{usr},u_1^{sys}, ...,u_{t-1}^{usr},u_{t-1}^{sys}, u_{t}^{usr})$$
where $P_a$ represents the action selection prompt, $I$ represents the product information, such as product description, price, etc.; $u_i^{usr}$ and $u_i^{sys}$ respectively represent the buyer's utterance and the agent's response.

After selecting the action to execute, the LLM needs to generate a corresponding response. However, LLMs are not naturally negotiation experts and tend to produce similar expressions when repeatedly executing the same bargaining actions. Therefore, in addition to defining relevant bargaining actions, we utilized large models to summarize a set of common bargaining language skills $S$ from real negotiation data, as detailed in Table~\ref{tab:script}. For each turn, we use a random selection of language skills $s_t \sim Random(S)$ to guide the LLM in generating persuasive language. Additionally, it is important to note that we focus on seller agents and we have omitted the definition of buyers' bargaining strategies.

\subsubsection{Bi-directional thinking}

In bargaining dialogues, the challenge lies not only in deciding the next action but also in the asymmetric and incomplete information game. Therefore, besides inferring one's actions based on the available information, it is also necessary to anticipate the actions the other party might take in the future to reflect on the rationality of one's own actions. We refer to this process as bi-directional thinking. We use bi-directional thinking to enable LLMs to better achieve bargaining objectives:
$$
a_t = LLM(P_a; P_b; I; u_{<=t})
$$
where $P_b$ is the bi-directional thinking prompt.

\subsection{Utterance Generator}

In order to positively influence the bargaining process, the utterances generated should comprehensively integrate the policy planner's decision on action and language skill. Specifically, if the action chosen is price-related such as COUNTER, we need to additionally generate a price based on the latest buyer's offer and seller's bottom price. 
% While there is much to explore in the price generator from an academic perspective, it is not in the range of this work.
% % 待定
To simplify this module, we propose a simple price sampler by truncated normalized distribution $p_t^{sys} \sim N(l_t, h_t)$. For each turn of the offer action, we define the price lower bound $l_t$ and upper bound $h_t$, and calculate the sampling centroid and delta. In conclusion, at generation stage, given action $a_t$ and language skill $s_t$, an LLM model is adopted to generate natural language. 
$$u_t^{sys} = LLM(P_u; I; u_{<=t}; a_t; s_t; p_t^{sys})$$
where $P_u$ represents the response prompt. 
% To ensure diversity, we adjust the temperature and topK parameters to produce multiple utterance candidates, and randomly choose one of them as result. 
\section{EXPERIMENTS}
\subsection{Data}

We collect real bargaining dialogues from Xianyu. Unlike CraigslistBargain~\cite{DBLP:conf/emnlp/HeCBL18}, which uses a limited number of annotators to simulate the bargaining environment, our dataset is sourced from actual buyer-seller conversations. We select the dialogues with bargaining behavior, and employ both GPT-4o0513~\cite{achiam2023gpt} and human annotators to process the data. After careful processing, we obtain 11,130 high-quality samples.
\setlength{\tabcolsep}{9pt}
\begin{table}[htbp]
  \centering
  \caption{Comparison of dataset statistics of Ours, CraigslistBargain~(CB), DealorNodeal~(DN), and SettlersOfCatan~(SoC).}
  % \vspace{-5pt}
  \label{tab:data}
    \begin{tabular}{lcccc}
    \toprule
          & Ours & CB & DN & SoC   \\
    \midrule
    \midrule
    \# of dialogues & 11,130 & 6,682 & 5,808 & 1,801  \\
    Avg \# of turns & 12.1 & 9.2 & 6.6 & 8.5  \\
    \# of categories & 1,492 & 6 &  - & -  \\
    \# of buyers/sellers & 11,130 & - & - & -     \\
    \bottomrule
    \end{tabular}%
  \label{tab:addlabel}%
    % \vspace{-10pt}
\end{table}%

Table~\ref{tab:data} shows the comparison of dataset statistics across datasets. With longer dialogues and more categories, our dataset has a richer corpus. Since each dialogue comes from real interactions buyer-seller interactions, our dataset effectively reflects the bargaining behaviors in real-world situations. Moreover, our dataset captures various reasons and tactics used by buyers and sellers to persuade each other during negotiations.
% , providing a robust foundation for developing and evaluating negotiation systems. 
% \subsection{MODEL}

% We choose GPT-4o0513 as the model for buyer bot.
% % ,  which supports up to 4096 tokens. 
% Meanwhile, we select qwen2-72B~\cite{yang2024qwen2} as the seller bot and policy planner. The price extractor module is obtained by fine-tuning based on the qwen2-1.5B model. For the parameters of the above models, we used the default configuration and did not make any special configuration.

\subsection{Metrics}

In line with previous research on bargaining dialogue, we focus on dialogue-level interactive evaluation. To this end, we employ the Average Turn(AT) and the Success Rate (SR) for the automatic assessment of goal achievement, and we use the Sale-to-List Ratio (SL\%) as a measure of goal achievement effectiveness in bargaining dialogues. Specifically, AT measures the efficiency of goal completion by calculating the average number of turns to achieve it, while SR assesses the effectiveness of goal completion by calculating the success rate within a predetermined maximum number of turns. SL\% is formalized as:

$$SL\%=max(0, min(1, \frac{deal\_price -seller\_lowest\_price}{seller\_price -seller\_lowest\_price }))$$
where a higher SL\% percentage indicates that the buyer gains more benefits from the transaction.

We conduct a human evaluation on 100 randomly sampled dialogues, where three annotators compare the baselines' generated responses with those from FishBargain.
\begin{figure}[h]
    \centering
    % 左图
    \begin{subfigure}[b]{0.45\textwidth} % 增加子图宽度
        \centering
        \includegraphics[width=\textwidth]{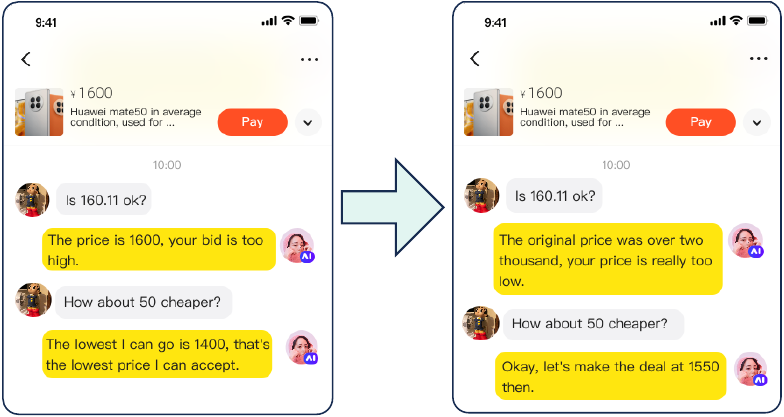}
        \caption{Comparison before and after adding the Price Extractor.}
        \label{fig:sub1}
    \end{subfigure}
    % \vspace{10pt}

    \begin{subfigure}[b]{0.45\textwidth}
        \centering
        \includegraphics[width=\textwidth]{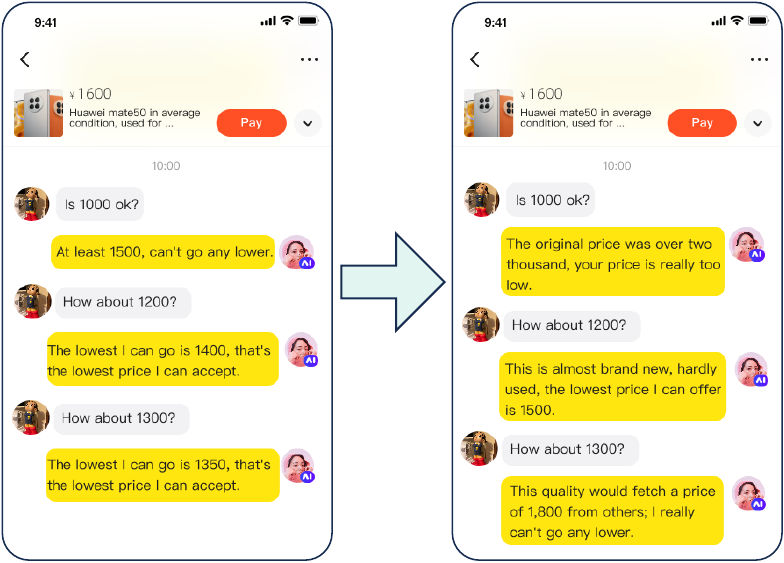}
        \caption{Comparison before and after adding the Policy Planner.}
        \label{fig:sub2}
    \end{subfigure}
    \vspace{-5pt}
    \caption{Illustration of FishBargain responses.}
    \label{fig:main}
    \vspace{-10pt}
\end{figure}
\subsection{Results}
To evaluate the proposed system, we use GPT-4o to judge transaction occurrences and prices at each turn $t$, with results shown in Table~\ref{tab:main_res}. For human evaluation, we assess four indicators: \textbf{Per}suasive, \textbf{Coh}erent, \textbf{Nat}ural, and \textbf{Ove}rall. As shown in Table~\ref{tab:hum}, FishBargain generally outperforms other agents in human evaluation, except for a higher win rate in Natural by +Price Extractor.
\setlength{\tabcolsep}{12pt}
\begin{table}[hbp]
  \centering
  \caption{Main results. Bold font denotes the best performance.}
  % \vspace{-5pt}
  \label{tab:main_res}
    \begin{tabular}{lccc}
    \toprule
    Model & AT$\downarrow$    & SR$\uparrow$    & SL\%$\uparrow$\\
    \midrule
    \midrule
    Baseline & 5.20  & 0.5530  & 0.0484  \\
    \midrule
    +  Price Extractor & 4.80  & 0.6820  & 0.1492  \\
    +  Action & 5.36  & 0.5610  & 0.1134  \\
    +  Language Skills   & 5.15  & 0.5955  & 0.0964  \\
    +  Bi-directional & 5.15  & 0.6075  & 0.1168  \\
    \midrule
    ALL   & \textbf{4.78}  & \textbf{0.6930}  & \textbf{0.1540}  \\
    \bottomrule
    \end{tabular}%
    \vspace{-10pt}
  \label{tab:addlabel}%
\end{table}%

\setlength{\tabcolsep}{2pt}
\begin{table}[htbp]
  \centering
  \caption{Human evaluation results.}
  \vspace{-5pt}
    \begin{tabular}{lrrrrrrrrrr}
    \toprule
    \multicolumn{1}{c}{FishBargain} & \multicolumn{2}{c}{Per.} & \multicolumn{2}{c}{Coh.} & \multicolumn{2}{c}{Nat.} & \multicolumn{2}{c}{Ove.}  \\
    \cmidrule(lr){2-3}\cmidrule(lr){4-5}\cmidrule(lr){6-7}\cmidrule(lr){8-9}  \multicolumn{1}{c}{vs.} & \multicolumn{1}{c}{Win} & \multicolumn{1}{c}{Loss} & \multicolumn{1}{c}{Win} & \multicolumn{1}{c}{Loss} & \multicolumn{1}{c}{Win} & \multicolumn{1}{c}{Loss} & \multicolumn{1}{c}{Win} & \multicolumn{1}{c}{Loss} \\
    \midrule
    \midrule
    Baseline     &  \textbf{98\%}     &    2\%   &    \textbf{100\%}  &     0\%  &    \textbf{92\%}   &    6\%   &   \textbf{100\%}   &            0\% &  \\
    +  Price Extractor    & \textbf{61\%}      &  28\%     &    \textbf{36\%}   &    14\%   &    30\%   &  \textbf{44\%}     &            \textbf{50\%}  &  44\%     &   \\
    +  Action &  \textbf{70\%}      &  30\%      &   \textbf{41\%}     &   37\%     &    \textbf{51\%}    &   30\%     &       \textbf{63\%}        &    37\%    &  \\
    +  Language Skills     &   \textbf{79\%}    &    20\%   &     \textbf{86\%}  &     10\%  &   \textbf{91\%}    &      5\% &     \textbf{76\%}  &          22\%    &  \\
    +  Bi-directional     &    \textbf{93\%}   &   5\%    &    \textbf{14\%}   &     1\%  &     \textbf{41\%}   &  4\%     &           \textbf{93\%}    &   4\%    &  \\
    \bottomrule
    \end{tabular}%
  \label{tab:hum}%
    \vspace{-10pt}
\end{table}%

\section{CONCLUSION}

% This paper presents a meaningful dialogue task: bargaining, derived from real user scenarios observed from online fleamarket platforms. Due to proactivity and quantitative aspects of this dialogue task, directly applying LLMs results in unsatisfactory outcomes. To deal with the deficiencies mentioned above, we introduce price extractor, policy planner and utterance generator. The three key modules collectively form the bargaining agent, named as FishBargain. FishBargain has offered a novel approach to enhance the user experience of individual sellers, and has been tested on Xianyu. This agent enables sellers with limited time investment and bargaining expertise to make more deals on the Xianyu platform.
This paper presents a meaningful dialogue task: bargaining, derived from real user scenarios observed from online fleamarket platforms.Directly applying LLMs to this task yields unsatisfactory results due to its proactive and quantitative nature. To address these issues, we propose a price extractor, policy planner, and utterance generator, forming the bargaining agent called FishBargain. FishBargain enhances the user experience for individual sellers on Xianyu, allowing those with limited time and bargaining skills to make more deals.

FishBargain just started as a project, demonstrating significant potential for more development and refinement. 
In the future, we will focus on more domain knowledge, better planner, and other aspects to continue optimizing the system.
% We have planned three possible future works. First, in order to generate more human-like utterances, Direct Preference Optimization(DPO) will be adopted to guide the training of generator. Second, more fleamarket related dialogue data will be utilized to infuse the LLM foundation models with domain-specific knowledge. Last, we will evaluate whether increasing Monte Carlo searches can improve the performance of the policy planner. 
In conclusion, FishBargain represents a substantial step forward in creating effective bargaining agents, and as we continue to improve the whole system, it will be more intelligent and provide better user experience.

\section{ETHICAL USE OF DATA}

% The dataset used for training and evaluating our agent was collected through human annotation. All data used in our research complies with ethical standards. All participants involved in the labeling process were provided with detailed information about the research work and the usage of annotated data. Measures were taken to anonymize all data collected to ensure data privacy. Moreover, during the test phase, all user data was stored in accordance with data protection regulations, and only authorized personnel could access the data for academic purposes.
The dataset for training and evaluating our agent was collected via human annotation. Participants in the labeling process were informed about the research and data usage. Data was anonymized to ensure privacy. During testing, user data was stored according to data protection regulations, with access restricted to authorized personnel for academic purposes.
\bibliographystyle{ACM-Reference-Format}
% \balance
\bibliography{sample-base}

\end{document}